\documentclass[epj]{webofc}
\usepackage[varg]{txfonts} 
\usepackage{slashed}
\wocname{EPJ Web of Conferences}
\woctitle{CONF12}
\begin{document}
\selectlanguage{english}
\title{OSEFT or how to go beyond hard thermal loops}
%
%

\author{Cristina Manuel\inst{1}\fnsep\thanks{\email{cmanuel@ice.cat}} \and
        Joan Soto\inst{2} \and
        Stephan Stetina\inst{3}
}

\institute{Instituto de Ciencias del Espacio (IEEC/CSIC) C. Can Magrans s.n., 08193 Cerdanyola del Vall\`es, Catalonia, Spain
\and
       Departament de Fisica Qu\`antica i Astrofisica
                   and Institut de Ci\`encies del Cosmos,
        Universitat de Barcelona,
        Mart\'\i $\,$ i Franqu\`es 1, 08028 Barcelona, Catalonia, Spain.
\and
           Institute for Nuclear Theory, University of Washington, Seattle, WA 98195, USA
}

\abstract{%
  We show that effective field theory techniques can be applied 
in the high temperature $T$ regime of plasmas to
improve the accuracy of the physics of the hard scales (or scales of order $T$), and as a by-product, also that of the soft scales
(or scales of order $gT$).
At leading order in the
coupling constant  the hard scales of the plasma can be viewed as
on-shell classical particles. Based on this observation, and without any
reference
to the state of the system, we derive an effective field
theory describing the quantum fluctuations around an on-shell fermion  with
energy $p$, described as a set of high dimension operators over the on-shell energy
$p$. When applied to systems close  to thermal equilibrium, where for most on-shell particles $p \sim T$, 
 we show that the
on-shell effective field theory (OSEFT) properly describes the  HTL photon
polarization tensor of QED, and its $1/T$ corrections.  For the soft scales the first non-vanishing power correction turns out to be a perturbative correction to
the HTL result.
}
\maketitle
\section{Introduction}
\label{intro}
A new effective field theory has been recently developed  that is aimed to improve many perturbative computations in  thermal field theory \cite{Manuel:2014dza,Manuel:2016wqs}. One of the main difficulties in thermal field theory comes from the fact that
even when the coupling constant $g$ is small, at high temperature $T$ the loop and coupling constant expansions do not coincide. However, in this regime there is a well-defined set of hierarchy of scales, 
the basic ingredient for the use of effective field theory (EFT) techniques. In the following we will explain the basic rationale and key ingredients of the on-shell effective field theory (OSEFT), summarizing the work presented in 
 \cite{Manuel:2016wqs}.
 
In the high temperature $T$ regime of 
QED or QCD there are one-loop thermal contributions, the so called hard thermal loops (HTLs) that are as relevant as the tree amplitudes for soft momenta of order
$gT$,  and  consequently have to be considered on the same footing as the tree level amplitudes
\cite{Braaten:1989mz,Frenkel:1989br}. Resummation of HTLs into effective propagators and vertices
is then mandatory for the soft scales. This program has been successfully used for the computation of different physical properties of hot QED and QCD plasmas,
but little is known on how the dynamics of the soft scales is perturbatively corrected in terms of the gauge coupling constant.

After the discovery of the existence of these HTL amplitudes it was soon realized that they could 
 be recovered from classical transport equations \cite{Blaizot:2001nr,Litim:2001db}. The hard degrees of freedom of the plasma, that is, those with momenta of the order $T$, can then be viewed as on-shell quasiparticles propagating in the plasma and interacting with the soft gauge fields. The OSEFT was initially developed to deduce quantum corrections to the classical tranport equations that could describe quantum  anomalous  effects, at high $T$ and in the presence of a chiral misbalance. However, the same OSEFT can be used to describe more accurately, beyond the classical trajectory approximation, the hard sector of the plasma. As the HTLs are obtained after integrating out the scales of order $T$, a more accurate description of the hard sector of the theory would also translate into a better description of the soft scales, as we will show  later on.

\section{The OSEFT: Lagrangian and Thermal  Propagators}
\label{sec-2}

The rationale and even the basic technical steps to deduce the OSEFT Lagrangian are essentially the same as those of other EFTs, such as 
HDET, NRQED, NRQCD, SCET, etc, which have been our source of inspiration (for a summary of those EFT see \cite{Brambilla:2004jw} and \cite{Hong:1998tn}).
 In any of those cases, one fixes a high energy scale (for example, the chemical potential
for HDET, or the mass for NRQED/NRQCD) and then defines some small fluctuations around the high energy scale. After integrating out the high energy modes, one is left with an EFT for the lower scales or quantum fluctuations. The resulting Lagrangian is organized in operators of increasing dimension over powers of
the high energy scale so that all the information on the high energy scales (beyond the explicit powers) is encoded in the matching coefficients of these
operators. The matching coefficients are obtained by enforcing the EFT to be equivalent to the fundamental theory at a given order
in the ratio of scales and/or in some small parameter, typically a coupling constant. 

We will only deal with the case of massless QED, although these  ideas can be generalized to massive QED, and also to QCD.
In our case, we will define as the high energy scale  the energy $p$ of the on-shell particle.
We will split the momentum of an almost on-shell fermion as
\begin{equation}
q^\mu = p v^\mu + k^\mu \ ,\label{eq:Qpart}
\end{equation}
where $v^\mu= (1, {\bf v})$ is a four-light-like velocity, and  $k^\mu$ is the residual momentum, {\it i.e}. the part of the momentum which makes $q^\mu$ off-shell.
 A similar decomposition of the momentum for almost on-shell antifermions can be done  as follows
\begin{equation}
q^{\mu}=-p\tilde{v}^{\mu}+k^{\mu}\,\label{eq:QAntiPart}
\end{equation}
where  $\tilde{v}^{\mu}=(1,-\bf{v})$~.

The Dirac field describing an almost on-shell propagation can be written factoring out its $p$-dependence
\begin{equation}
\psi_{\bf v}=e^{-ipv\cdot x}\left(P_{ v}\chi_{v}(x)+P_{\tilde v}H_{\tilde v}^{(1)}(x)\right)+e^{ip\tilde{v}\cdot x}\left(P_{\tilde v}\xi_{\tilde v}(x)+P_v H_{v}^{(2)}(x)\right) \ ,
\label{eq:Fields}
\end{equation}
where 
%
\begin{equation}
P_v  =  \frac{1}{2}\gamma\cdot v\,\gamma_{0}\,, \qquad \,
P_{\tilde v}  =  \frac{1}{2}\gamma\cdot\tilde{v}\,\gamma_{0}.
\end{equation}
are the particle and antiparticle projectors, respectively. The fields describing the particle and antiparticle quantum fluctuations, 
$\chi_{v}(x)$ and $\xi_{\tilde v}(x)$, respectively, contain soft momenta only ($k_\mu \ll p$), whereas $H_{v}^{(1)}(x)$ and $H_{v}^{(2)}(x)$ contain generic off-shell  momenta. After integrating out the latter, one can get en effective Lagrangian for the former, expressed as
\begin{equation}
\label{general-L}
 \mathcal{L}=\sum_{p, {\bf v}}\mathcal{L}_{p, {\bf v}}
 \end{equation}
and which can be expanded in powers of $1/p$  (for notations and conventions, please see ~\cite{Manuel:2016wqs}).
Thus for the particle quantum fluctuations one gets

\begin{eqnarray}
 \label{Lan-0}
 \mathcal{L}^{(0)}_{p, v}\ &= & \chi_{v}^{\dagger} \left(i\, v\cdot D\,\right)\chi_{v} \ , \\
 \label{Lan-1}
  \mathcal{L}^{(1)}_{p, v}\ &= &
   - \frac {1}{2 p}\chi_{v}^{\dagger} \,\left( D_{\perp}^{2} - \frac e2 \sigma^{\mu \nu}_\perp F_{\mu \nu} \right)\chi_{v} \ , 
 \end{eqnarray}
for the first orders in the expansion. 
 The interaction terms generated at order
 $1/p^{2}$, and higher, contain temporal derivatives. In order to simplify the computations
 of different Feynman diagrams at this and higher orders it is convenient to perform  local
 field redefinitions, such that only temporal derivatives acting on the fermionic fields appear in $\mathcal{L}^{(0)}_{p, v}$. 
Thus, one gets
\begin{eqnarray}
  \label{Lan-2}
\mathcal{L}^{(2)}_{p, v} &= & \frac{1}{8p^{2}} \chi_{v}^{\prime\dagger} \Big(\left[ \slashed{D}_{\perp}\,,\,\left[i\tilde{v}\cdot D\,,\,\slashed{D}_{\perp}\right]\right] -
\left\{ ( \slashed{D}_{\perp})^{2},\,\left(iv\cdot D-i\tilde{v}\cdot D\right)\right\}  \Big) \chi_{v}^{\prime} \ ,\\
\label{Lan-3}
\mathcal{L}^{(3)}_{p, v} & = & \frac{1}{8p^{3}} \chi_{v}^{\prime\prime\dagger}\left\{ \slashed{D}_{\perp}^{4}+\left[\slashed{D}_{\perp},i\tilde{v}\cdot D\right]^{2}-(iv\cdot D-i\tilde{v}\cdot D)\slashed{D}_{\perp}^{2}(iv\cdot D-i\tilde{v}\cdot D)\right\} \chi_{v}^{\prime\prime}\label{eq:L3}\\
 & + & \frac{1}{8p^{3}} \chi_{v}^{\prime\prime\dagger}\left\{ (iv\cdot D-i\tilde{v}\cdot D)\slashed{D}_{\perp}\,\left[i\tilde{v}\cdot D,\slashed{D}_{\perp}\right]-\left[i\tilde{v}\cdot D,\slashed{D}_{\perp}\right]\slashed{D}_{\perp}(iv\cdot D-i\tilde{v}\cdot D)\right\} \chi_{v}^{\prime\prime} \ ,\nonumber 
\end{eqnarray}
with no dependence on temporal derivatives.

The Lagrangian for the antiparticle fluctuations $\xi_{\tilde v}$ is easily obtained after performing
the changes $p \leftrightarrow - p$ and $v^\mu \leftrightarrow {\tilde v}^\mu$ in the Lagrangian for the particle fluctuations.

For computations of properties at finite temperature we will use the real time formalism, in the Keldysh representation.
One can deduce the thermal propagators in two different ways. Either starting from the thermal propagator in QED, projecting it over both the particle and antiparticle sectors, using the splitting of momenta in 
Eqs.~(\ref{eq:Qpart}) and (\ref{eq:QAntiPart}), respectively. Or alternatively, deduce them from the Lagrangian of the OSEFT just described, but realizing that  $p$ (or $-p$) acts as
a chemical potential for the particle fluctuations (minus chemical potential for the antiparticles). This last observation can be substantiated  if we write the Hamiltonian of the theory in terms of the
fields $\chi_v$ and $\xi_{\tilde v}$.

More explicitly, the retarded, advanced and symmetric particle propagators obtained in the Keldysh formalism read
\begin{eqnarray}
\label{RAF-propa}
S^{R/A}(k) & = &  \frac{ P_v \gamma_0   }{k_0  \pm i \epsilon - f({\bf k})} ,\\
S^{S}(k)  & = & P_v \gamma_0 \left( - 2\pi i \delta( k_0 - f({\bf k})) \left( 1 - 2n_f(p +k_0)  \right) \right)  \, .
\end{eqnarray}
The expansion of f({\bf k}) at order $n$  will be denoted as $f^{(n)}({\bf k})$. At lowest order
\begin{equation}
f^{(0)}({\bf k}) =   k_\parallel \ ,
\end{equation}
and we have defined $ k_\parallel = {\bf k} \cdot {\bf v}$, while 
\begin{equation}
f^{(1)}({\bf k}) = k_\parallel + \frac{{\bf k}_\perp^2}{2 p} \ , \qquad
 f^{(2)}({\bf k}) = k_\parallel + \frac{{\bf k}_\perp^2}{2 p} - \frac{k_\parallel {\bf k}_\perp^2}{2 p^2 } \ ,
 \label{displaw-2}
  \end{equation}
which can be deduced from the OSEFT Lagrangians written above.

The propagators of the OSEFT should  also be expanded  for large $p$, keeping in mind that the thermal distribution function $n_f$ should also be Taylor expanded, and that
a term proportional to $d^n n_f/dp^n$ contributes at the same order as  one $ \sim 1/p^n$.

 The OSEFT mainly will help us in
organizing the contributions to different Green's functions in powers of $1/p$. By considering the vertex interactions, deduced from the 
OSEFT Lagrangians, and the propagators obtained at a given order in $1/p^n$, one can write
down the different Feynman diagrams contributing at a given order in $1/p^n$.
Because in a thermal bath most of
the on-shell quasiparticles have energies of the order $p \sim T$, ultimately this expansion will help us in computing
$1/T^n$ corrections to the different amplitudes.
This will be shown in the following section with an explicit example.

\section{Retarded Photon Polarization Tensor at One-Loop}

We have  computed the one-loop retarded photon polarization tensor within the OSEFT up to order $n=3$ in the $1/p$ expansion.
Here we summarize the main results, and give some relevant details that help in performing these computations.

\begin{figure}[h]
\centering
\includegraphics[width=6.5cm,clip]{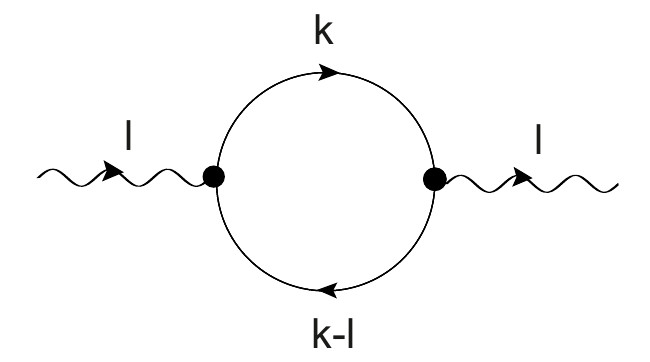}
\caption{Bubble Feynman diagram}
\label{fig-1}       
\end{figure}

\begin{figure}[h]
\centering
\includegraphics[width=5.0cm,clip]{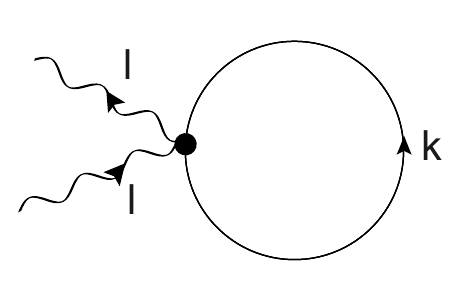}
\caption{Tadpole Feynman diagram: they arise in the OSEFT to give account of photon-particle/antiparticle interactions mediated by an off-shell antiparticle/particle.}
\label{fig-2}       
\end{figure}

There are two set of different topological distinct Feynman diagrams: the bubble (see figure ~\ref{fig-1}) and the  tadpole (see figure ~\ref{fig-2}). The latter arises to give account of particle-photon interactions mediated by an off-shell antiparticle (or vice versa for the antiparticle-photon interactions). The two topologies are required at every order in the $1/p$ expansion to preserve the
gauge symmetry and to fulfil the Ward identity obeyed by the photon polarization tensor.

The computation of loop diagrams in the OSEFT involves the evaluation of integrals over the residual momentum $k$, which in principle require the introduction of a cutoff  recalling
that $ k_\mu \ll p$. A prescription of the sum over on-shell energies and different velocity directions
 that appears in Eq.~(\ref{general-L}) and in the loop integrals should also be given. It turns out convenient, in order to avoid dealing with
cutoff dependent results, to revert the change of variables of Eqs.~(\ref{eq:Qpart}) and (\ref{eq:QAntiPart}) before the integrals are performed. Then one uses the identification
\begin{equation}
 \sum_{p, {\bf v}}\int\frac{d^{3}{\bf k}}{(2\pi)^{3}} \equiv \int\frac{d^{3}{\bf q}}{(2\pi)^{3}} \ ,
\label{def}
\end{equation} 
and derives from Eq.~(\ref{eq:Qpart})
\begin{eqnarray}
\label{p-back}
p &=& q - k_{\parallel,{\bf q}} + \frac{{\bf k}_{\perp,{\bf q}}^2 }{2 q} +{\cal O}(\frac{1}{q^2}) \ ,\\
\label{v-back}
{\bf v} & = & {  \bf{\hat q}} -  \frac{{\bf k}_{\perp,{\bf q}} }{q}  - \frac{  {\bf \hat q} {\bf k}_{\perp,{\bf q}}^2 + 2  k_{\parallel,{\bf q}} {\bf k}_{\perp,{\bf q}}}{2 q^2}+{\cal O}(\frac{1}{q^3})
\ , \\
\label{n-back}
n_f(p) &= & n_f(q) +  \frac{d n_f}{dq} \left( -  k_\parallel^{\bf q}  + \frac{{\bf k}_{\perp,{\bf q}}^2 }{2 q} \right) + \frac 12 \frac{d^2 n_f}{dq^2}   k_{\parallel, {\bf q}}^2 +
{\cal O}(\frac{1}{q^3}) \ ,
\end{eqnarray}
where we have defined $k_{\parallel,{\bf q}} \equiv {\bf k} \cdot { \bf{\hat q}}$, where ${  \bf{\hat q}}=\frac{\bf q}{q}$, $q=\vert {\bf q}\vert$, 
and ${\bf k}_{\perp,{\bf q}} \equiv{\bf k} - {  \bf{\hat q}} k_{\parallel,{\bf q}}$. All the Feynman diagrams are then finally expressed as an integral over ${\bf q}$.
Then a consistency check of the OSEFT is that, after reverting the change of variables, the
$k$-dependence of the different amplitudes at a given order  should vanish exactly in the final result. This  cancellation of the $k$-dependence typically involves different
Feynman diagrams that may appear at different orders in $n$.

From the basic set of rules already explained, it is not difficult to see that one easily recovers the HTL of the photon polarization tensor from the OSEFT at $n=1$, after considering the contribution of both the bubble and tadpole diagrams for both particle and antiparticle fluctuations. Thus we arrive at the following 
 longitudinal and transverse components of the polarization tensor
\begin{eqnarray}
\label{pipiL}
\Pi^L_{(1)} (l_0, {\bf l}) & = & m^2 _{D} \left( \frac{l_0}{2|{\bf
l}|} \left(
 \,{\rm ln\,}\left|{\frac{l_0+|{\bf l}|}{l_0-|{\bf l}|}}\right| 
-i \pi \, \Theta(|{\bf l}|^2 -l_0^2) \right)
-1  \right) \ , \\
 \Pi^T_{(1)} (l_0, {\bf l}) & = &- m^2 _{D} \, \frac{l_0^2}{2 |{\bf
l}|^2} \left[ 1 + \frac12 \left( \frac{|{\bf l}|}{l_0} -
\frac{l_0}{|{\bf l}|} \right) \, \left( {\rm ln\,} \left|{\frac{l_0+
|{\bf l}|}{l_0-|{\bf l}|}}\right| -i \pi \, \Theta(|{\bf l}|^2 -l_0^2)
 \right) \, \right] \ .
 \label{pipiT}
\end{eqnarray}
which exactly matches the HTL result. Here $m^2_D = e^2T^2/6$ is the Debye mass squared.

The contribution to the photon polarization tensor computed at $n=2$ vanishes exactly,  meaning that there
is no correction proportional to $\alpha T $, where $\alpha$ is the electromagnetic fine structure constant, to the HTL result. 

The first correction to the HTL polarization tensor appears at order $n=3$. The bubble diagrams are straightforwardly computed. But
there is a subtlety in the computation of the tadpole diagrams at order $n=3$. The UV logarithmic divergences which appear at this order are taken care of, as usual, with dimensional regularization, with $d= 3 + \epsilon$, but an additional prescription is needed for their computation, which otherwise, becomes ambiguous. This is due to the fact that the vertices in 
 Eq.~(\ref{eq:L3})  which appear in the tadpole diagrams contain high powers of derivative couplings. Two different prescriptions for their computation were given in  
 \cite{Manuel:2016wqs}.

The final result is matched to a  computation of the one-loop polarization tensor in QED, expanded for large internal loop momentum, that we carried out also to check the consistency of the approach.
Local counterterms have then to be added to the OSEFT Lagrangian to remove the UV divergences, but also to match the result of QED
\begin{equation}
{\cal L}_{c.t.}=-\frac{Z(\alpha , \epsilon) C(\alpha ,\mu)}{2}F_{0i}F^{0i}-\frac{Z'(\alpha , \epsilon) C'(\alpha ,\mu)}{4}F_{ij}F^{ij}
\label{c.t.} \ ,
\end{equation}
where $Z$ and $Z'$ stand for the counterterms, while $C$ and $C'$ stand for the matching coefficients. Their explicit values depend on the prescription used to compute the tadpoles,
and can be found in  \cite{Manuel:2016wqs}.

The final result, after considering both the tadpole and bubble diagrams of both particle and antiparticle fluctuations, depends on the renormalization scale $\mu$. 
Using the MS  renormalization scheme, and choosing  ${\mu}=\frac {\sqrt{\pi}}{2} T e^{-1 -\gamma/2}$, one then reaches to the following longitudinal and transverse components
of the polarization tensor at this order
\begin{eqnarray}
\label{final-Long}
\Pi^{L}_{(3)} (l_{0},{\bf l}) &= &\frac{\alpha}{  \pi }\left[{\bf l}^{2}- \frac 13 l_{0}^{2}+ \frac 16 \frac{l_{0}}{|{\bf l}|}\left(l_{0}^{2}-3{\bf l}^{2}\right)  \left(
 \,{\rm ln\,}\left|{\frac{l_0+|{\bf l}|}{l_0-|{\bf l}|}}\right| 
-i \pi \, \Theta(|{\bf l}|^2 -l_0^2) \right)
 \right] \ , \\
 \Pi^{T}_{(3)} (l_{0},{\bf l}) &= & \frac{2 \alpha}{\pi } \left[ \frac 49 l_{0}^{2} -\frac {43}{90} {\bf l}^{2}  + \frac 16\frac{l_{0}^{4}}{{\bf l}^{2}}- \frac {1}{12} \frac{l_{0}^{3}}{{|\bf l|}^{3}}
 \left(l_{0}^{2}+2{\bf l}^{2}-3\frac{{l}^{4}}{l_{0}^{2}}\right)  \right.  \nonumber \\  
& \times& \left.   \left(
 \,{\rm ln\,}\left|{\frac{l_0+|{\bf l}|}{l_0-|{\bf l}|}}\right| 
-i \pi \, \Theta(|{\bf l}|^2 -l_0^2) \right)  \right] \ .
\label{final}
\end{eqnarray}

After fixing the renormalization scale, the above correction does not depend on $T$, even if it arises   due to the thermal fluctuations in the plasma. While the HTL polarization tensor is proportional to $\alpha T^2$, the above result should be understood as arising as a correction to the HTL result behaving as $ \sim \alpha T^2 \frac{ l^2}{T^2} \sim \alpha l^2$. It is only at this order, $n=3$, that there is no  $T$-dependence in a pure thermal generated effect.

Note also that the above contribution to the polarization tensor represents a perturbative correction to the HTL. For soft momentum $ l \sim e T$, then 
 $\Pi_{(3)}/\Pi_{(1)} \sim \alpha$.  However, this is not the only $\alpha$ correction to the HTL physics, as there are two-loop diagrams, with hard loop momenta, that contribute 
 as $\alpha^2 T^2$, and thus for soft momenta they have to be considered on equal footing as the one-loop piece $\Pi_{(3)}$. This is just another example of the fact that 
 the loop and gauge coupling constant expansions do not coincide in thermal field theory, as mentioned in the Introduction. We expect to report in the near future on how the OSEFT might also help in obtaining those two-loop contributions.

\section{Discussion}

We have devised an EFT  approach to compute properties mainly dominated by on-shell degrees of freedom. While the formalism is completely general and does not require the
presence of a thermal bath, its applications in thermal field theory are particularly interesting. This is so because in this case we know that some physical properties of hot plasmas are dominated
by the contribution of on-shell degrees of freedom. We have focused our interest here in showing that the  OSEFT allows one to obtain corrections to the HTL photon polarization tensor. It would be very interesting to analyze other Feynman diagrams beyond one-loop with the same techniques, to finally
obtain the full perturbative correction to the HTL polarization tensor. It might be particularly interesting to generalize all the approach to QCD.
The OSEFT may also help in computing perturbative corrections to transport effects, as well as quantum corrections to classical transport equations. 
We hope to report soon on these efforts.

\vspace{1cm}

{\bf Acknowledgements:} We have been supported by the CPAN  CSD2007-00042 Consolider--Ingenio 2010 program, and the and FPA2013-43425-P
projects (Spain). J.S. also acknowledges support from the 2014-SGR-104 grant (Catalonia) and the FPA2013-4657 project (Spain).S.S. has been partially supported by the Schroedinger Fellowship of the FWF, project no. J3639 (Austria).

\end{document}